\documentclass{styles/osa-article}

\journal{oe}
\begin{document}
\title{A wide-range wavelength-tunable photon-pair source for characterizing single-photon detectors}

\author{Lijiong~Shen,\authormark{1} Jianwei~Lee,\authormark{1} Antony Winata~Hartanto,\authormark{2} Pengkian~Tan,\authormark{1} and Christian~Kurtsiefer \authormark{1,2,*}}

\address{\authormark{1}Centre for Quantum Technologies, National University of Singapore, 3 Science Drive 2, Singapore 117543, Singapore\\
\authormark{2}Department of Physics, National University of Singapore, 2 Science Drive 3, Singapore 117551, Singapore\\}
\email{\authormark{*}christian.kurtsiefer@gmail.com} %% email address is required

\begin{abstract} 
The temporal response of single-photon detectors is usually obtained by measuring their impulse response to short-pulsed laser sources. 
In this work, we present an alternative approach using time-correlated photon pairs generated in spontaneous parametric down-conversion (SPDC).
By measuring the cross-correlation between the detection times recorded with an unknown and a reference photodetector, the temporal response function of the unknown detector can be extracted.
Changing the critical phase-matching conditions of the SPDC process provides a wavelength-tunable source of photon pairs. 
We demonstrate a continuous wavelength-tunability from 526\,nm to 661\,nm for one photon of the pair, 
and 1050\,nm to 1760\,nm for the other photon.
The source allows, in principle, to access an even wider wavelength range by simply changing the pump laser of the SPDC-based source.
As an initial demonstration,
we characterize single photon avalance detectors sensitive to the two distinct
wavelength bands, one based on Silicon, the other based on Indim Gallium Arsenide.
\end{abstract}

\section{Introduction} 
Characterizing the temporal response function of single-photon detectors 
is crucial in time-resolved measurements, e.g. determining the lifetime of fluorescence markers~\cite{Becker2005}, 
characterizing the spontaneous decay of single-photon emitters~\cite{Stevens2006} and the photon statistics of
astronomical sources~\cite{Tan2016} and measuring the joint spectral of photon-pair sources~\cite{zielnicki2018joint}, so that the timing uncertainty contributed by the detection process can be taken into account.
Typically, the temporal response of a detector is obtained from the arrival time distribution of photons collected from a pulsed laser. 
In this work, we present an alternative approach that leverages on the tight
timing correlation~\cite{HongOuMandel1987} of photon pairs generated in
spontaneous parametric down-conversion~\cite{Klyshko70,Burnham70} (SPDC): 
the coincidence signature corresponding to the detection of two photons of the
same pair is used to infer the temporal response function of the
photodetectors. Compared to a pulsed laser, a SPDC source
is easier to align, and is
wavelength-tunable by changing the critical phase-matching condition of the
SPDC process~\cite{Harris1969}. In addition, one can address two
wavelength bands with the same source by choosing a non-degenerate phase
matching condition. 

For an initial demonstration, we generate photon pairs with a tunable
wavelength range over 100\,nm in the visible band, and over 700\,nm in the telecommunication band
-- a tunability at least comparable to existing femtosecond pulsed lasers -- 
and use it to characterize both Silicon (Si-APDs) and Indium Gallium Arsenide (InGaAs-APDs) avalanche photodiodes.
In particular, we characterize timing the behaviour of a fast commercial Si-APD (Micro
Photon Devices PD-050-CTC-FC) over a continuous wavelength range, for which we
previously assumed an approximately uniform temporal response of the detector
in the wavelength range from 570\,nm to 810\,nm~\cite{Tan2016}. With the
measurement reported in this work, we observe a significant variation of the
timing jitter even on a relatively small wavelength interval of $\approx$10\,nm.
A better knowledge of the timing response of this particular Si APD contributes to a better
understanding of coherence properties of light in such experiments.
Similarly, better characterization of the timing response over a wide wavelength
range helps to better model fluorescence measurements regularly carried out
with such detectors~\cite{Becker2005}.

\section{Correlated photon pair source}\label{sec:experiment}
\begin{figure} %[h]
%\centering\includegraphics[width=.8\linewidth]{figures/layout.pdf}
\centering\includegraphics[width=.8\linewidth]{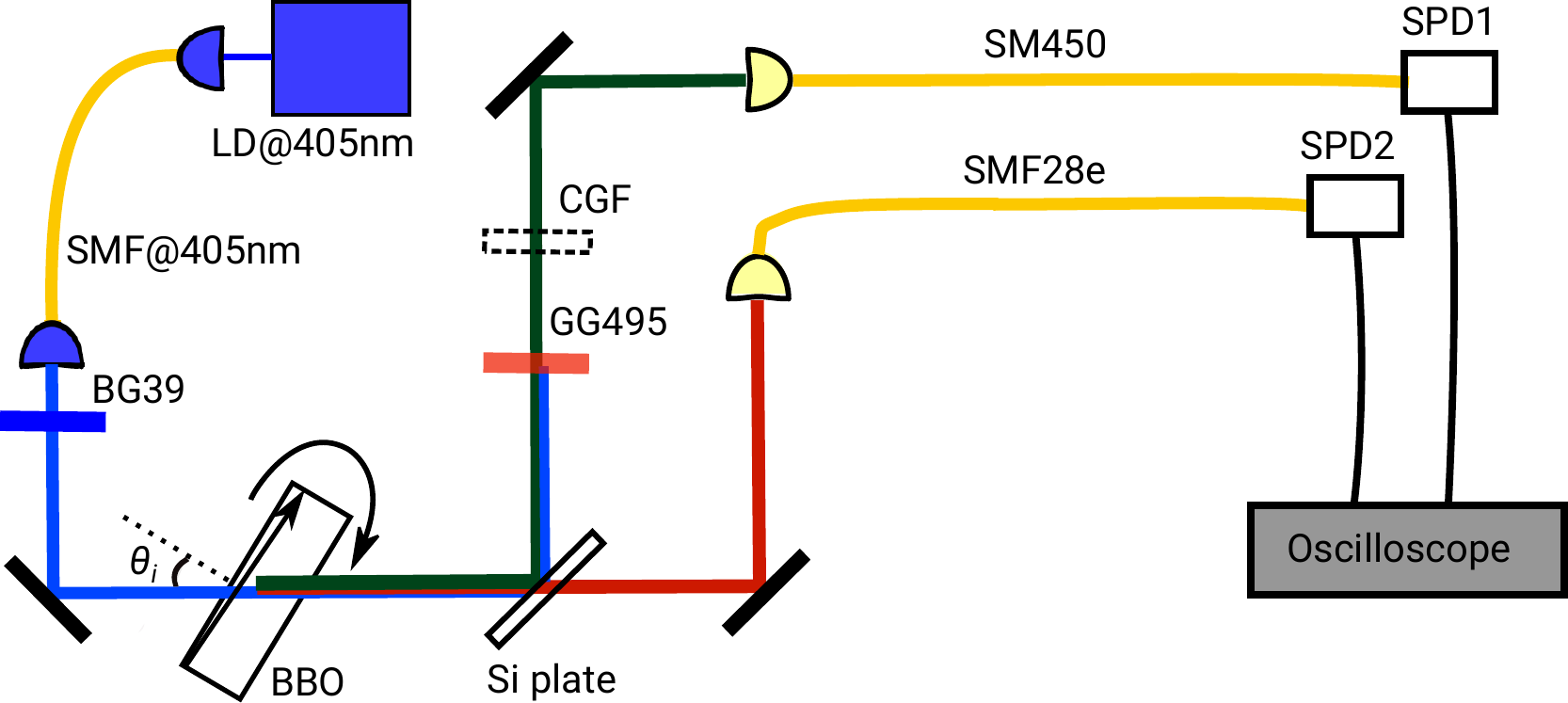}
\caption{
    Wavelength-tunable photon pair source based on Type-II SPDC. 
    The critical phase-matching condition is changed by
    varying the angle of incidence  $\theta_i$ of the pump at the crystal,
    in order to generate photon pairs at the desired wavelength in the visible and telecommunications band. 
    A Silicon (Si) plate separates the photons in each pair. 
    Tight timing correlations between photons in each pair, and a characterized detector SPD2, allow measuring the jitter of a single-photon detector (SPD1). 
    A calibrated color glass filter (CGF) can be inserted to infer the
    wavelength of the photons sent to SPD1 using a transmission measurement.
    LD: laser diode, 
    BBO: $\beta$-Barium Borate, 
    SMF: single-mode fiber, 
    GG495, BG39: color glass filters. 
     }
     \label{fig:setup}
\end{figure}
The basic configuration of the spontaneous parametric down conversion source
is shown in Fig.~\ref{fig:setup}.
The output of a laser diode (central wavelength $\lambda_p=405$\,nm, output power 10\,mW)
is coupled to a single-mode optical fiber for spatial mode filtering, and
focused to a Gaussian beam waist of 70\,$\mu$m into a 2\,mm thick
$\beta$-Barium Borate crystal  as the nonlinear optical element, cut for
Type-II phase matching ($\theta_0=43.6^{\circ}$, $\phi=30^{\circ}$).

For this cut, SPDC generates photon pairs in the visible and telecommunications band, respectively. 
We collect the photons in a collinear geometry, with collection modes (beam waists $\approx 50\,\mu$m) defined by two single-mode fibers:
one fiber (SMF450: single mode from 488\,nm to 633\,nm) collects signal photons and delivers them to the single-photon detector SPD1,
while the other fiber (standard SMF28e, single transverse mode from 1260\,nm to 1625\,nm) collects idler photons and delivers them to SPD2.
The signal and idler photons are separated to their respective fibers using a
100\,$\mu$m-thick, polished Silicon (Si) plate as a dichroic element.
The plate acts as a longpass filter (cut-off wavelength $\approx 1.05\,\mu$m),
transmitting only the idler photons while reflecting approximately half of the signal photons. 

To suppress uncorrelated visible and infrared photons detected by our SPDs, 
we insert both a blue color glass bandpass filter (BG39) in the pump path,
attenuating parasitic emission from the pump laser diode and broadband
fluorescence from the mode cleaning fiber,  
and a green color glass longpass filter (GG495) in the path of the idler
photons to suppress pump light at SPD1. For the idler path, the silicon
dichroic is sufficient.
\begin{figure}[!t]
%\centering\includegraphics[width=.7\linewidth]{figures/wavelength_vs_theta/angle_wavelength2.eps}
\centering\includegraphics[width=.7\linewidth]{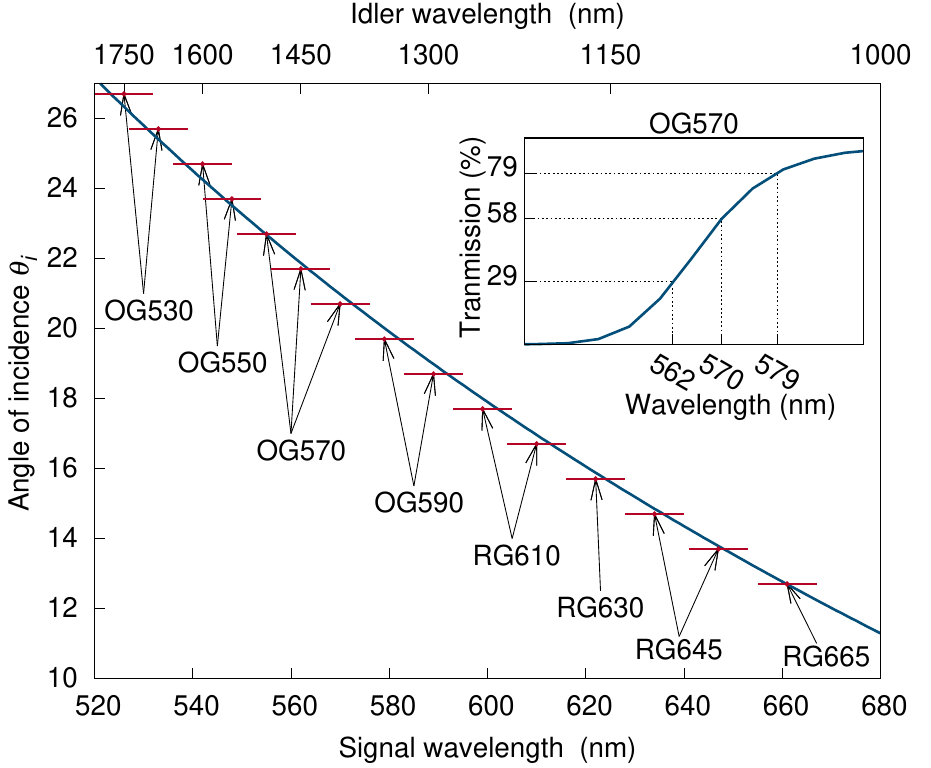}
  \caption{
  Signal ($\lambda_s$) and idler ($\lambda_i$) wavelength dependence on the
  angle of incidence $\theta_i$ of the pump beam  at the $\beta$-BBO
  crystal, 
  produced in Type-II SPDC. 
  We obtain $\lambda_s$ by measuring the transmission of signal photons through a set of calibrated color glass filters (OGs and RGs). 
  $\lambda_i$ is calculated from $\lambda_s$ by the law of energy conservation. 
  Error bars: uncertainty of the central transmission wavelength of the filters. 
  Solid line: model predicting the output wavelengths based on energy and momentum conservation laws governing SPDC. 
  Inset: the transmission-wavelength calibration curve of color glass filter OG570 used to experimentally determine three values of $\lambda_s$.
     }
  \label{fig:lambda}
\end{figure}
To tune the wavelength of down-converted photons, we 
change the critical phase-matching condition of the SPDC process by
varying the angle of incidence $\theta_i$ of the pump beam  at the crystal~\cite{sutherland,Robert1}.
Figure~\ref{fig:lambda} (red dots) shows the signal and idler wavelengths,
$\lambda_s$ and $\lambda_i$, measured for our source for $\theta =
12.7^{\circ}$ to $26.7^{\circ}$. To measure the signal wavelength $\lambda_s$,
we insert different standardized color glass longpass filters (CGF in Fig.~\ref{fig:setup}) for
different angles $\theta_i$, and measure the transmission of the signal photons
in order to infer their wavelength. The inset of Fig.~\ref{fig:lambda}
shows an example where a filter OG570 is used to infer $\lambda_s$ close to
the cut-off wavelength of the filter. The corresponding idler wavelength
is calculated through energy conservation in SPDC,
$\lambda_i^{-1}=\lambda_p^{-1}-\lambda_s^{-1}$. Our measured SPDC wavelengths
can be well described by a numerical phase matching model based on optical
dispersion properties of BBO~\cite{Eimerl1987,Nikogosyan1991} (blue line).

This simple pair source provides photons in a
wavelength range of $\lambda_s=526$\,nm to 661\,nm and $\lambda_i=1050$\,nm to 1760\,nm, comparable with existing dye and solid-state femtosecond pulsed lasers~\cite{Duarte1990,Sorokin2005}.
In the following section, we demonstrate how the tight timing correlations of
each photon pair can be utilized to characterize the temporal response of
single photon detectors.

\section{Characterizing the temporal response of single-photon
  detectors}\label{sec:theory}
The time response function $f(t)$ of a single photon detector characterizes
the distribution of signal events at a time $t$ after a photon (of a
sufficiently short duration) is absorbed by a detector. It characterizes the
physical mechanism that converts a single excitation into a macroscopic
signal, and can be measured e.g. recording the average response to attenuated
optical pulses from a femtosecond laser~\cite{Antia2007}. In this paper, we use the timing correlation
in a photon pair, which emerges at an unpredictable point in time. This
requires two single photon detectors registering a photon. As the photon pair is
correlated on a time scale of femtoseconds, and the relevant time scales for
detector reponses is orders of magitudes larger, the correlation
function $c_{12}(\Delta t)$ of time differences $\Delta t$ between the
macroscopic photodetector signals is a convolution of the two detector response functions,
\begin{equation}\label{eq:g2}
  c_{12}(\Delta t) = N(f_1 * f_2)(\Delta t) = N\int f_1(t)\,f_2(\Delta t-t) dt\,,
\end{equation}
where $N$ the total number of recorded coincidence events.
Obtaining the detector response function $f_1(t)$ from a measured
correlation function $c_{12}(\Delta t)$ requires the known response function
$f_2(t)$ of a reference detector. For a device under test,  $f_1(t)$ can then be
either reconstructed by fitting a $c_{12}(\Delta t)$ in
Eqn.~\ref{eq:g2} with a reasonable model for $f_1(t; P)$ (with a parameter set
$P$) to a measured correlation function, or obtained from it via deconvolution.

To measure $c_{12}(\Delta t)$, we evaluate the detection time at single photon
detector SPD1 by recording the analog detector signal with an
oscilloscope, and interpolating the time it crosses a threshold of around
half the average signal height with respect to a trigger event caused by a
signal of single photon detector SPD2. The histogram of all time differences
$\Delta t$ for many pair events then is a good represenation of $c_{12}(\Delta t)$.

\section{Reference detector characterization}\label{sec:refchar}

We use a superconducting nanowire detector (SNSPD) with a design wavelength at
1550\,nm as the reference detector SPD2, because a SNSPD has an intrinsic
wide-band sensitivity and fast temporal response. To determine its response
function $f_2(t)$, we measure the correlation function $c_{12}$ from photon pairs with two
detectors of the same model (Single Quantum SSPD-1550Ag).

\begin{figure}[t]
%\centering\includegraphics[width=0.7\columnwidth]{figures/nanowire/readout2.pdf}
\centering\includegraphics[width=0.7\columnwidth]{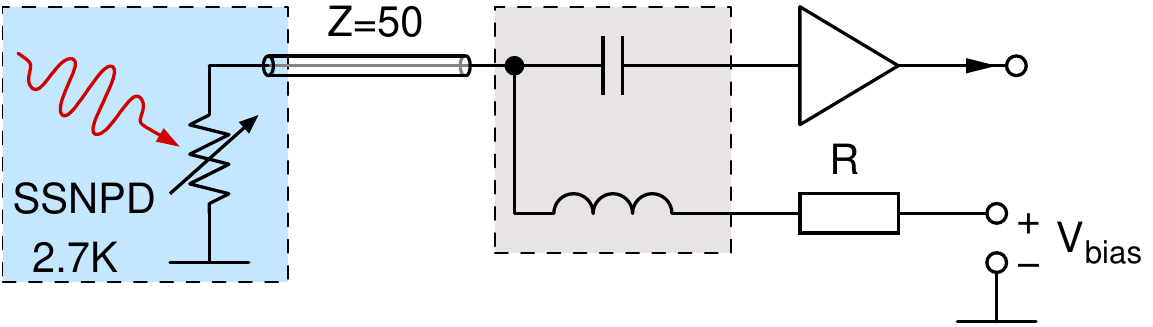}
\caption{
  Biasing and readout circuit for the superconducting nanowire single-photon detector (SNSPD). 
  The SNSPD is current-biased using a constant voltage source and a series resistor $R$.
  When a photon is absorbed by the SNSPD, it changes temporarily from a
  superconducting to a conducting state.
  The resulting current change reaches a signal amplifier, which provides the photodetection signal.
  }
  \label{fig:nanowire}
\end{figure}

Figure~\ref{fig:nanowire} shows the biasing and readout circuit of a single SNSPD.
The SNSPD is kept at a temperature of 2.7\,K in a cryostat, and is
current-biased using a constant voltage source ($V_{\text{bias}}=1.75$\,V) and
a series resistor ($R= 100\,\text{k}\Omega$) through a bias-tee at room
temperature. The signal gets further amplified by 40\,dB at room
temperature to a peak amplitude of about 350\,mV.

\begin{figure}
%\centering\includegraphics[width=.65\linewidth]{figures/nanowire/setup_g2.pdf}
\centering\includegraphics[width=.65\linewidth]{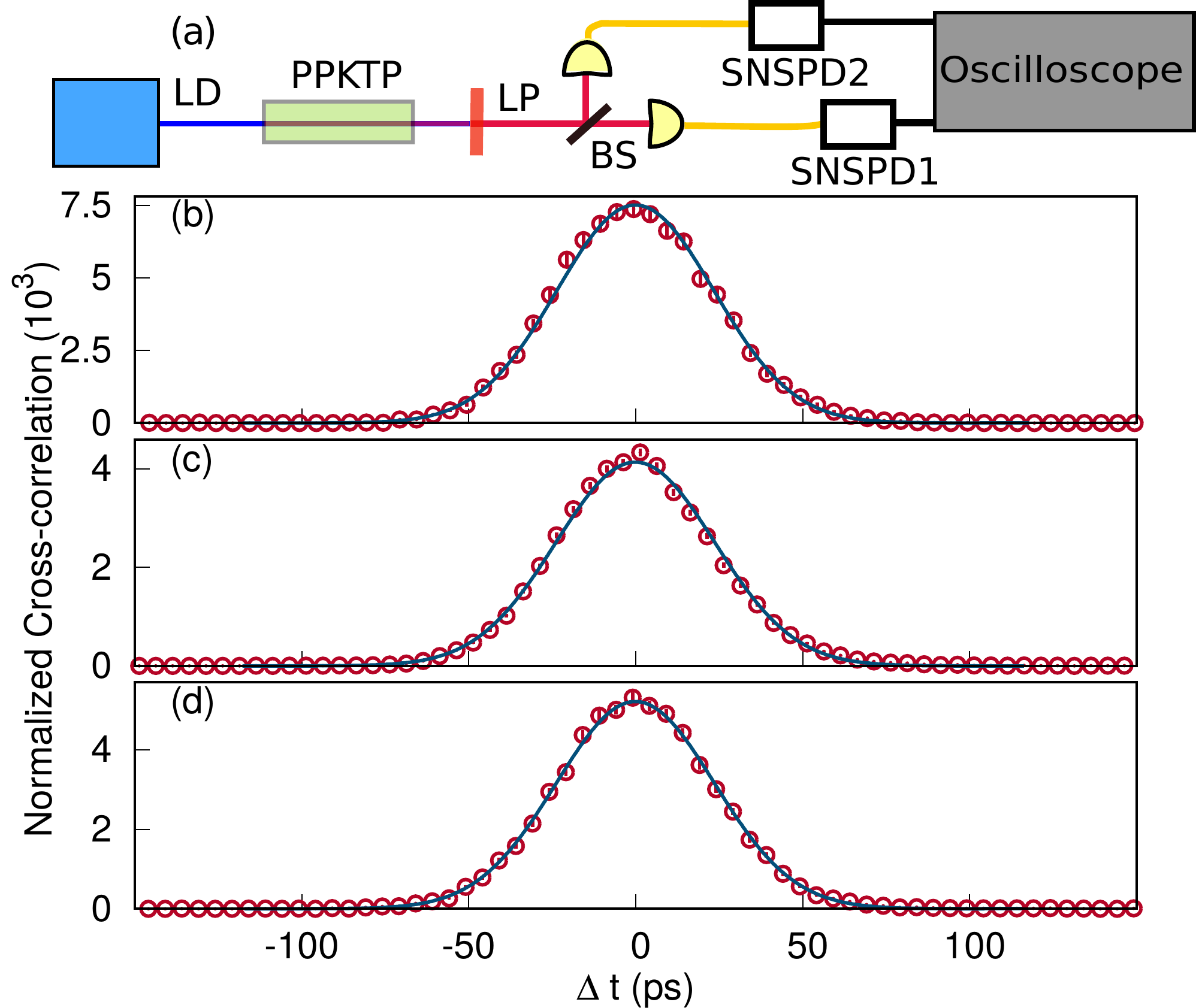}
\caption{\label{fig:jitter_snspd}
  (a) Simplified schematic of the PPKTP-based photon-pair source.
  We generate orthogonally-polarized, degenerate photon pairs at 810nm using Type-II SPDC, 
  by pumping a PPKTP crystal (9.55$\,\mu$m poling period) with a 405 nm laser beam. 
  The photon pairs are separated by a polarizing beam-splitter and fibre-coupled to two SNSPDs. 
  By changing the pump wavelength to 532 nm, the same crystal generates 810 nm and 1550 nm photon pairs in a Type-0 SPDC process. 
  We use a Si plate as a beam splitter to separate the non-degenerate photon pairs. BS: beam splitter, LD: laser diode, LP: longpass filter, PPKTP: Periodically Poled Potassium titanyl phosphate crystal. 
  Cross-correlation of photodetection times registered by two SNSPDs detecting
  (b) degenerate 810\,nm photon pairs, and (c) non-degenerate photon pairs at 810\,nm and 1550\,nm from the PPKTP-based source.
  (d) Cross-correlation of photodetection times of non-degenerate photon pairs
  at 548\,nm and 1550\,nm from the BBO-based source in Fig.~\ref{fig:setup}.
  }
\end{figure}

We first expose both detectors to photons at a wavelength of 810\,nm using a degenerate PPKTP-based photon pair source pumped with a 405\,nm laser diode (Fig.~\ref{fig:jitter_snspd}\,(a)).
The choice of using this source instead of the BBO-based source shown in
Fig.~\ref{fig:setup} was borne out of convenience rather than from any
limitation in our BBO-based source described before, as
the PPKTP-based type-II SPDC source was readily available~\cite{Shen2018a}.
Figure~\ref{fig:jitter_snspd}\,(b) shows the cross-correlation $c_{12}(\Delta
t)$
for the two SNSPDs, normalized to background coincidences (red dots).

The histogram closely follows a Gaussian distribution (blue line) with
standard deviation $\sigma_{12}= 23.6(1)$\,ps. This suggests that the two
responses $f_1(t), f_2(t)$ are also Gaussian distributions, and
Eqn.~\ref{eq:g2} can be simplified to
\begin{equation}\label{eq:g2_gau1}
   c_{12}(\Delta t) = N G(\sigma_{12}, \Delta t) + C_0 = N G(\sigma_1, \Delta t)*G(\sigma_1, \Delta t) + C_0\,,
\end{equation}
where $N$ is the total number of correlated photon pairs detected, 
$G(\sigma, \Delta t)=e^{-\Delta t^2/(2\sigma^2)}/\sqrt{2\pi\sigma^2}$ is a
normalized Gaussian distribution,
and $C_0$ is associated with the accidental coincidence rate. The standard
distribution of the correlation is then simply related to those of the individual
detectors by $\sigma_{12}^2=\sigma_1^2+\sigma_2^2$. Assuming the same
response for both detectors, we can infer
at a wavelength of 810\,nm, corresponding to a the full-width at half-maximum
(FWHM) of 39.2(2)\,ps.

Next, we calibrate the SNSPD at 1550\,nm using photon pairs at 810\,nm and
1550\,nm generated from the same PPKTP-based SPDC source pumped with a
$532\,$nm laser diode [Fig.~\ref{fig:jitter_snspd}\,(a)]. The non-degenerate
photon pairs are separated by a Si plate as a dicroic element.  
Figure~\ref{fig:jitter_snspd}\,(c) shows the cross-correlation (red dots) of the photodetection times at the two SNSPDs, and a fit of a Gaussian distribution (blue line) with a standard deviation $\sigma_{12,810/1550}=23.8(2)$\,ps.
With $\sigma_{1,810}=16.7(1)$\,ps obtained at 810\,nm previously, 
we obtain
$\sigma_{2,1550}=\sqrt{\sigma_{12,810/1550}^2-\sigma_{1,810}^2}$
  resulting in a timing jitter of 39.9(6)\,ps (FWHM) of the SNSPD at 1550\,nm.

Finally, to determine the temporal response function of a SNSPD at 548\,nm, we
used the BBO-based pair source [Fig.~\ref{fig:setup}] to prepare non-degenerate photon pairs at 548\,nm and 1550\,nm.
Figure~\ref{fig:jitter_snspd}\,(d) shows the cross-correlation obtained with
our detectors. The fit to a Gaussian distribution (blue line) leads to a
standard deviation $\sigma_{12, 548/1500}=23.7(1)$\,ps.
With the same argument as before, and using $\sigma_{1,1500}=16.9(2)$\,ps,  we
obtain a timing jitter of 38.9(7)\,ps (FWHM) at 548\,nm.
So in summary, the timing jitter of the SNSPD shows no statistically
significant dependency on the wavelength in our measurements. 

The timing jitter partually originates from the threshold detection mechanism:
for a photodetection signal $V(t)$, the timing uncertainty $\sigma_{t}$ for
crossing a threshold, contributed by the electrical noise $\sigma_{V}$, is
given by $\sigma_{t; \text{noise}} = \sigma_{V} / (dV/dt)$ at the threshold~\cite{bertolini1968, Wu2019}. 
For our SNSPDs, we estimate $\sigma_{t;\text{noise}} \approx 15$\,ps,
corresponding to a contribution of about 35\,ps to the timing jitter of the
combined SNSPD and electronic readout system, i.e., we are dominated by
this electrical noise. 
The jitter of the oscilloscope is claimed to be a few ps, which suggests that
the intrinsic jitter of these SNSPDs is about $10-20$\,ps (FWHM)~\cite{Korzh2020}. 

In the following section, we use the standard deviation $\sigma_2$ obtained at these wavelengths to define the temporal response function of the reference detector $f_2 = G(\sigma_2)$ in Eqn.~\ref{eq:g2},
and use the method outlined in Sec.~\ref{sec:theory} to characterize $f_1$ of an unknown detector.

\section{Avalanche photodetector characterization}
\begin{figure}[t]
%\centering\includegraphics[width=.7\linewidth]{figures/mpd/combined.eps}
\centering\includegraphics[width=.7\linewidth]{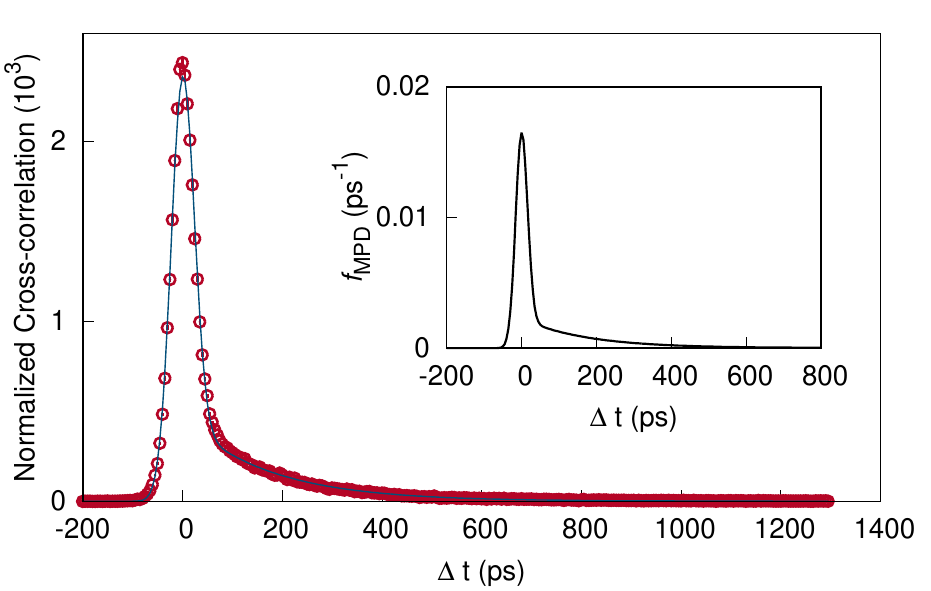}
  \caption{\label{fig:fit_mpd}
  Cross-correlation between photodetection times at a Si-APD and a characterized SNSPD, normalized to background coincidences $g^{(2)}(\Delta t)$. 
  The detectors were illuminated by a non-degenerate (555\,nm, 1500\,nm) photon pair source.
  By fitting the data (red dots) to a model (blue line) obtained by convolving the individual temporal response model of both detectors,
  we are able to extract parameters describing the temporal response of the Si-APD (inset).
  $\Delta t$: photodetection time difference.
     }
\end{figure}
First, we characterize the temporal response function $f_{\text{Si}}$ of a thin Silicon avalanche photodiode (Si-APD) from Micro Photon Devices (PD-050-CTC-FC).
Although thin Si-APDs have been characterized in previous works at
a few discrete wavelengths~\cite{Lacaita1989,Giudice2007,Cova1987a}, 
there has yet been a characterization performed over a continuous wavelength range. 

Following Refs.~\cite{Tan2016,Hernandez2017}, we describe the temporal response function with a heuristic model 
\begin{equation}\label{eq:mpd_mod}
f_{\text{Si}}(\Delta t) = A\,G(\sigma, \Delta t-\mu) + B\,G(\sigma, \Delta t-\mu)*e^{-(\Delta t - \mu)/\tau}\,,
\end{equation}
a combination of a Gaussian component of mean $\mu$ and standard deviation $\sigma$, and an exponential term with a characteristic decay constant $\tau$. 
The weights of each distribution are described by $A$ and $B$.
The Gaussian component is associated with an avalanche that occurs due to the absorption of a photon in the depletion region. 
The exponential component, convoluted with a Gaussian distribution, is associated with an avalanche that is initiated by a photoelectron that diffused into the depletion region produced by photon absorption elsewhere.

We characterize the Si-APD over a wavelength range from $\lambda_1$ = 542\,nm to 647\,nm in steps of about 10\,nm. 
The photon wavelength is tuned by rotating the crystal, changing the angle of incidence  $\theta_i$ of the pump from $13.7{^\circ}$ to $24.7{^\circ}$, in steps of $1{^\circ}$.
For each $\theta_i$, we
obtain the cross-correlation $c_{12}(\Delta t)$ similarly as in section~\ref{sec:refchar}. 
Figure~\ref{fig:fit_mpd} (red dots) shows $g^{(2)}$, the cross-correlation normalized to background coincidences, obtained when signal and idler wavelengths are $\lambda_1=555$\,nm and $\lambda_2=1500$\,nm, respectively.
For every $(\theta_i,\lambda_1,\lambda_2)$,
we deduce $f_{\text{Si}}$
by fitting the measured $c_{12}$ to the model in Eqn.~\ref{eq:g2}
with $f_1 = f_{\text{Si}}$,
and $f_2$ a Gaussian distribution with full-width at half-maximum (39.9\,ps) corresponding to the SNSPD jitter at 1550\,nm. 
For the SNSPD, we assume that its jitter remains constant over the
wavelength range $\lambda_2=1082$\,nm to 1602\,nm, motivated by the
observation that it does not differ significantly for $\lambda_2=810$\,nm and 1550\,nm.
The fit results in parameters $\sigma$ and $\tau$ which characterize
$f_{\text{Si}}$ at the corresponding wavelength
$\lambda_1$. Figure~\ref{fig:fit_mpd} (inset) shows $f_{\text{Si}}(\Delta t)$ for $\lambda_1=555$\,nm.

\begin{figure}
%\centering\includegraphics[width=.7\linewidth]{figures/mpd/combined_mpd.eps}
\centering\includegraphics[width=.7\linewidth]{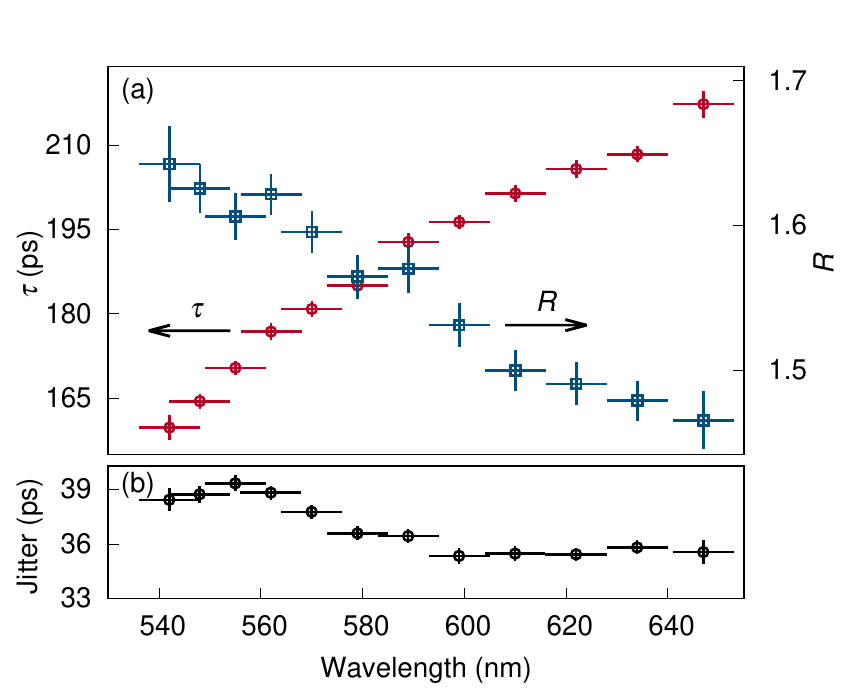}
  \caption{\label{fig:ratio} (a)
  Si-APD temporal response characteristics:
  ratio $R$ between the Gaussian and the exponential component of the temporal
  response function (red), and characteristic decay constant $\tau$ of the
  exponential component(blue). (b) Measured timing jitter of a Si-APD as a
  function of wavelength.
  Horizontal error bars: uncertainty of the cut-off wavelength of the color glass filter used to measure photon wavelength.
  Vertical error bars: fit error of $R$, $\tau$, and timing jitter.
  }
\end{figure}

Two figures of merit are of interest for characterizing the thin Si APD:
the duration  $\tau$ of the exponential tail, and the ratio $R$ between the coincidences attributed to the Gaussian component to those attributed to the exponential component,
\begin{equation}
R = \frac{\int_{-\infty}^{\infty}A\,G(\sigma, \Delta
  t-\mu)}{\int_{-\infty}^{\infty}B\,G(\sigma, \Delta t-\mu)*e^{-(\Delta t - \mu)/\tau}} = \frac{A}{B\tau}\,.
\end{equation}
Both values determine if the full-width at half-maximum (FWHM),
a value typically quoted for the detector jitter,
serves as a good figure of merit for the temporal response of a detector.
For example, the jitter of a detector with $R \ll 0.5$ and $\sigma \ll \tau$,
is better described by $\tau$ than the FWHM of the temporal response
function.  

Figure~\ref{fig:ratio}\,(a) shows that $R$ reduces while $\tau$ increases with
increasing wavelength. The detector jitter (FWHM) is shown in Figure~\ref{fig:ratio}\,(b).
The observation that $\tau$ changes significantly with wavelength is
especially revelant for fluorescence lifetime measurements, where 
the exponential tail in the temporal response function can be
easily misattributed to fluorescence when the detector is not characterized at the wavelength of interest~\cite{Becker2005}. 

Next, we characterize the temporal response function $f_{\text{InGaAs}}$ of an
InGaAs avalanche photodiode (S-Fifteen Instruments ISPD1) which is
sensitive in the telecommunication band.
\begin{figure}
%\centering\includegraphics[width=.7\linewidth]{figures/ingaas/fitted_ingaas.eps}
\centering\includegraphics[width=.7\linewidth]{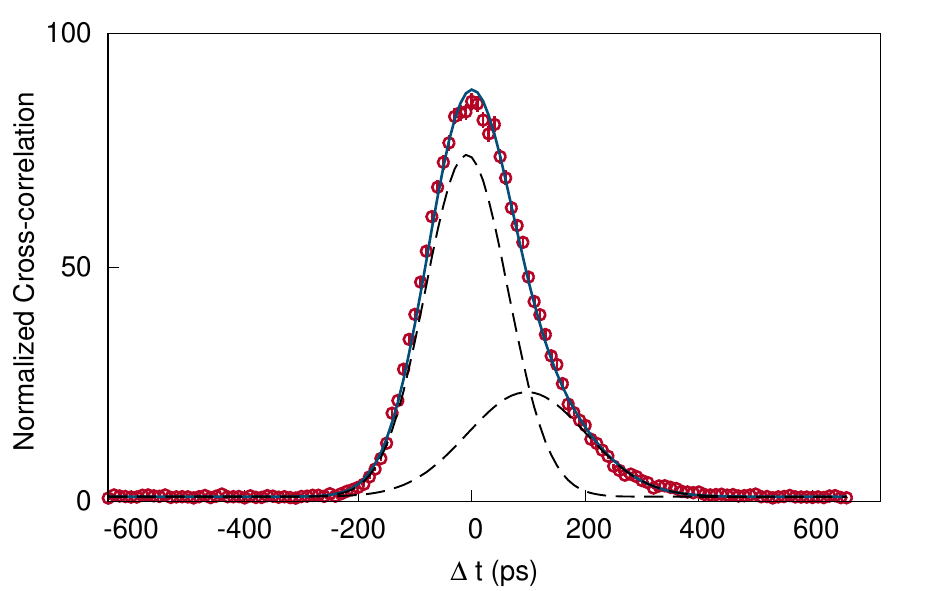}
  \caption{\label{fig:ingaas} Cross-correlation function normalized to background coincidences $g^{(2)}(\Delta t)$ of the InGaAs-APD and the reference SNSPD. 
  The cross-correlation approximates the InGaAs-APD temporal response well since the latter is much slower than the SNSPD.
  We fit the measured $g^{(2)}(\Delta t)$ (red dots) with a model consisting
  of two Gaussian distributions (solid line) with an overall width of 196\,ps
  (FWHM).
  Dashed lines: individual Gaussian components,   $\Delta t$: time difference between the photodetection times.
  }
\end{figure}
We extract $f_{\text{InGaAs}}$ by measuring the cross-correlation $c_{12}$ of the detection times between the InGaAs-APD and our reference SNSPD.
We note that since the expected jitter of the InGaAs-APD ($\approx 200\,$ps) is significantly larger than that of the SNSPD ($\approx 40\,$ps), 
$f_{\text{InGaAs}}$ is well-approximated by $c_{12}$.

Again, we fit $c_{12}(\Delta t)$ to a heuristic model~\cite{Tosi2011}, here comprising of a linear combination of two Gaussian distributions
\begin{align}\label{eqn:two_gau}
c_{12}(\Delta t) & \approx Nf_{\text{InGaAs}}(\Delta t)+C_0 = N[A\,G(\mu_1,\sigma_1,\Delta t) + B\,G(\mu_2,\sigma_2, \Delta t)]+C_0\,,
\end{align}
where $A$ and $B$ are the weights of each distribution, $\mu_{1}$ ($\mu_{2}$) and $\sigma_1$ ($\sigma_2$)  is the mean and standard deviation characterizing the Gaussion distribution G, and $C_0$ is associated with the accidental coincidence rate.
Figure~$\ref{fig:ingaas}$ shows the measured cross-correlation $c_{12}$ (red dots) 
and the fit result (blue line) when 
the InGaAs-APD detected photons with a wavelength of 1200\,nm.

\begin{figure}[t]
\begin{center}
\includegraphics[width=.65\linewidth]{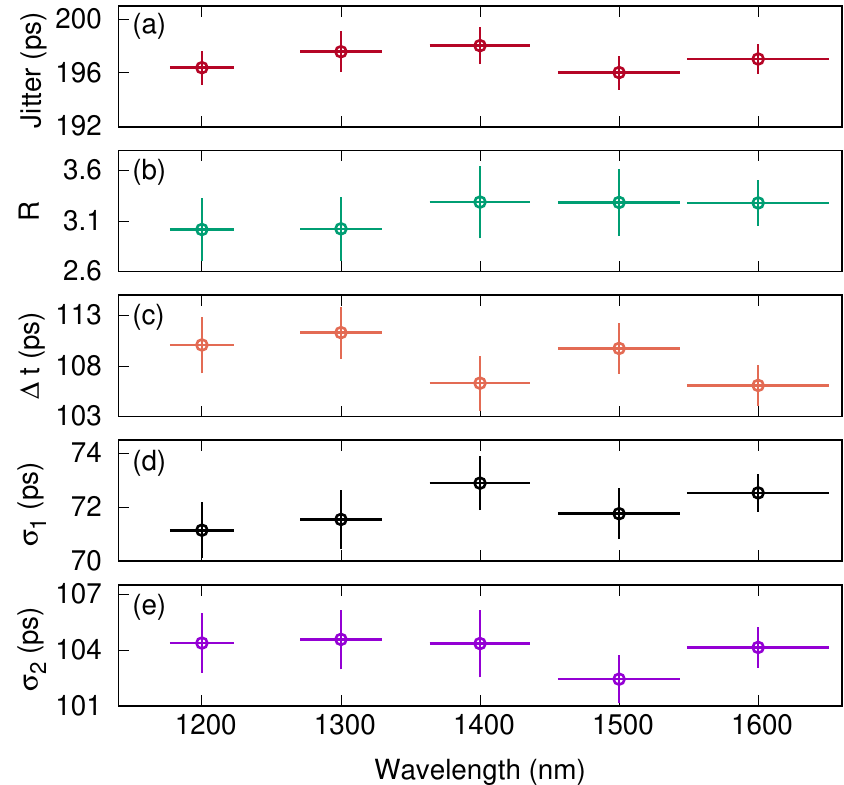}
  \caption{\label{fig:fwhm_ingaas} 
   Parameters describing the temporal response function $f_{\text{InGaAs}}$ of
   an InGaAs-APD, measured over a wide wavelength range for a linear
   combination of two Gaussian distributions to model $f_{\text{InGaAs}}$ (Eqn.~\ref{eqn:two_gau}).
   The parameters are extracted by fitting the measured temporal response to this model:
   (a) timing jitter,  
   (b) the weight ratio $R$ of the Gaussian distributions, 
   (c) the temporal separation $\mu_1-\mu_2$ between the Gaussian distributions, 
   (d) and (e) the standard deviations $\sigma_1$ and $\sigma_2$ of the Gaussian distributions.
  }
\end{center}
\end{figure}

We tune the wavelength of the photons sent to the InGaAs-APD from 1200\,nm to 1600\,nm in steps of 100\,nm,
and obtain $c_{12}$ for each wavelength. 
Figure~\ref{fig:fwhm_ingaas} shows the parameters describing the temporal
response of the InGaAs-APD: its jitter, the ratio $R = A/B$ of the two
Gaussian distributions contributing to $f_{\text{InGaAs}}$, the temporal
separation between the two Gaussian distributions ($\mu_1-\mu_2$), and the
standard deviation of the two Gaussian distributions
($\sigma_1$,$\sigma_2$). We find no significant variation of any parameter over the entire wavelength range.

\section{Conclusion}
We have presented a widely-tunable, non-degenerate photon-pair source that produces signal photons in the visible band, and idler photons in the telecommunications band.
With the source, we demonstrate how the tight-timing correlations within each photon pair can be utilized to characterize single-photon detectors.
This is achieved by measuring the cross-correlation of the detection times registered by the device-under-test (DUT), and a reference detector -- an SNSPD, which has a relatively low and constant jitter over the wavelength range of interest.
By taking into account the jitter introduced by the reference detector, we are able to extract the temporal response function of the DUT.
As the source is based on SPDC in a BBO crystal, its output wavelengths are continuously tunable by varying the angle of incidence of the pump at the crystal. 
We experimentally demonstrated wavelength-tunability of over 100\,nm in the visible band, and over 700\,nm in the telecommunications band -- a similar tunability compared to existing femtosecond pulsed laser systems. 

With our source, we measured the temporal response functions of two single-photon detectors, an Si-APD and an InGaAs-APD, over a continuous wavelength range centered at the visible and telecommunications band, respectively.
For the InGaAs-APD, we observed no significant variation of its jitter over a wide wavelength range.
For the Si-APD, we observed that the exponential component of its temporal response increases with wavelength.
This observation emphasizes the need for an accurate accounting of Si-APD
jitter in precision measurements, e.g. characterizing fluorescence markers at
the wavelength of interest~\cite{Becker2005}, or measuring the photon statistics of narrowband astronomical sources~\cite{Tan2016}.
\section*{Funding}
This research was supported by the National Research Foundation (NRF) 
Singapore (Grant: QEP-P1), and through the Research Centres of 
Excellence programme supported by NRF Singapore and the Ministry of Education,
Singapore. 

% \bibliographystyle{osajnl}
% \newpage
% \bibliography{abbo}

\end{document}